IAC-12-A2.5.11

# THE UNITED NATIONS HUMAN SPACE TECHNOLOGY INITIATIVE (HSTI) SCIENCE ACTIVITIES

**Aimin Niu**
United Nations Office for Outer Space Affairs, United Nations Office at Vienna, Austria,
aimin.niu@unoosa.org

**Mika Ochiai, Hans Haubold, Takao Doi**
United Nations Office for Outer Space Affairs, United Nations Office at Vienna, Austria
mika.ochiai@unoosa.org, hans.haubold@unoosa.org, takao.doi@unoosa.org

The United Nations Human Space Technology Initiative (HSTI) aims at promoting international cooperation in human spaceflight and space exploration-related activities; creating awareness among countries on the benefits of utilizing human space technology and its applications; and building capacity in microgravity education and research. HSTI has been conducting various scientific activities to promote microgravity education and research. The primary science activity is called "Zero-gravity Instrument Distribution Project", in which one-axis clinostats will be distributed worldwide. The distribution project will provide unique opportunities for students and researchers to observe the growth of indigenous plants in their countries in a simulated microgravity condition and is expected to create a huge dataset of plant species with their responses to gravity.

## I. INTRODUCTION

The United Nations Programme on Space Applications was established in 1971 [1] with the aim of assisting countries which lacked the human, technical and financial resources to fully utilize the benefits of space technology [2]. The Programme has been implemented by the United Nation Office for Outer Space Affairs (UNOOSA), the Secretariat of the United Nations Committee on the Peaceful Uses of Outer Space (COPUOS).

From 1971 to 2010, the Programme conducted 271 activities such as expert meetings, seminars, workshops, and international conferences in 67 different countries with a total number of 18,151 participants [3]. Five regional centres for space science and technology education, affiliated to the United Nations were established in response to the second UNISPACE (UNISPACE'82) [4]. Currently, the Programme covers the following wide range of space related thematic areas: environmental monitoring, natural resource management, satellite communications for tele-education and telemedicine applications, disaster risk reduction, development of capabilities in the use of global navigation satellite systems (GNSS), the Basic Space Science Initiative, space law, climate change, the Basic Space Technology Initiative, and the Human Space Technology Initiative [5].

The Human Space Technology Initiative (HSTI) was launched by UNOOSA in 2010 as the latest initiative under the framework of the Programme on Space Applications [6]. HSTI builds on the relevant recommendations related to human spaceflight and exploration contained in the report of the third United Nations Conference on the Exploration and Peaceful Uses of Outer Space (UNISPACE III) [7]. HSTI aims at promoting international cooperation in human spaceflight and space exploration-related activities, creating awareness among countries on the benefits of utilizing human space technology and its applications, and building capacity in microgravity education and research. In order to create awareness on the benefits of human space technology and microgravity science, HSTI has been conducting science outreach activities and is currently preparing to launch a new capacity-building project specifically focused on microgravity life science [8].

A microgravity environment provides excellent research, applications, and educational opportunities in the field of life science, material and physical science, Earth observation, and space science. In microgravity environments, scientists can observe and investigate new physical phenomena and processes that are normally masked by the effects of Earth's gravity on the ground. The results of experiments in microgravity have contributed to a better understanding of fundamental physics and of gravity influences on fluid phenomena and biological processes, including physiological adaptations of the human body to the new environment of space.

Space platforms such as Apollo, Skylab, Salyut, Mir, Soyuz, Spacelab, Space Shuttles, the International Space Station (ISS), Shenzhou, and Tiangong have provided ideal opportunities for microgravity science experiments in space. However, current understanding of the effects of gravity in physics, material science, and life science is still very limited.

From a physical point of view, any object in free fall experiences microgravity [9]. Based on this principle, ground-based facilities such as drop tubes, drop towers, parabolic flights, and sounding rockets were developed to provide reduced gravity for limited duration with different gravity levels ranging from $10^{-2}$-$10^{-6}$g [10]. Other ground-based facilities such as clinostats, slow rotating rooms, and bed test clinics were also developed to generate simulated reduced gravity, mainly for microgravity research in life science [11]. Small drop tubes and clinostats can also provide excellent opportunities for students to learn in a classroom about the effects of gravity.

## II. SCIENCE OUTREACH ACTIVITIES

In order to raise awareness of the benefits of human space technology and microgravity science, specifically in developing countries, HSTI has organized several outreach activities, including the Outreach Seminar on the International Space Station in February 2011 and the Expert Meeting on Human Space Technology in November 2011.

### II.I Outreach Seminar on the International Space Station on 8 February 2011

The Outreach Seminar on the International Space Station was held in Vienna, Austria, on 8 February 2011, during the 48th session of the Scientific and Technical Subcommittee of COPUOS [12]. The seminar was organized by the UNOOSA in close cooperation with the ISS partners. Representatives from 17 different countries participated in the seminar.

The ISS is 110 meters in length and 70 meters in width. It weighs more than 360 metric tons, and the power generation with eight solar arrays is approximately 80 kilowatts. It provides a unique environment for microgravity science research and space applications and is an ideal tool to support educational activities as well as a gateway for human exploration of the solar system.

In order to participate in ISS utilization, a non-ISS partner country needs to define the research it wants to carry out and the expected outcomes as well as to estimate national readiness to participate in terms of available national professional expertise and funding.

There is also a need to train younger generations on the importance of developing capacity-building activities in this area. Educational materials made available by ISS partners in print and on the WWW greatly contribute to boosting that effort. It was also emphasized in the seminar that the real key would be building local capacity for the advancement of science and technology in support of national priority needs and development goals in human space flight benefits.

The seminar observed that HSTI could be a meaningful mechanism for creating awareness on the potential of the ISS and the research conducted on the ISS among countries, regions, and potential users which have up to this point not been involved in such activities, and thus, HSTI would contribute to capacity building in microgravity science and technology education in the world.

### II.II Expert Meeting on Human Space Technology on 14-18 November 2011

The United Nations/Malaysia Expert Meeting on Human Space Technology was held in Malaysia from 14 to 18 November 2011 [13]. Hosted by the Institute of Space Science of the National University of Malaysia and co-organized by UNOOSA and the ISS partner agencies, more than 120 experts from around the world participated in the meeting.

On-orbit space facilities such as the ISS can provide an ideal microgravity environment for research to better understand fundamental questions in science and to develop new technology for benefits on Earth. Additionally, such space facilities are regarded as suitable platforms to develop and verify technologies for long duration space exploration.

During the meeting, microgravity simulators such as clinostats, drop tubes, drop towers and parabolic flights were suggested as examples for ground-based microgravity research whose utilisation should be encouraged. It was proposed that countries establish national microgravity centres which could significantly contribute to the infrastructure and capacity building in microgravity science. A solid ground-based research programme using microgravity and hypergravity facilities was considered essential to facilitate spaceflight experimentation.

The importance of international cooperation in microgravity research was emphasized. Non-space-faring countries were encouraged to seek cooperation with space-faring countries through individual scientific collaborations, multinational institutional agreements or by establishing national or regional expert centres. Such initiatives can promote capacity building for independent national space science and technology programmes. Countries should also be encouraged to explore commercial opportunities in their effort to build capacities in human space technology.

The following 10 recommendations were formulated at the end of the meeting:

(a) The Human Space Technology Initiative should take action to create awareness among stakeholders, including decision makers in the public and private sectors, researchers and students, of the social and economic potential of space science and technologies and to initiate outreach activities;

(b) The Initiative should identify and inform Member States about space-related research opportunities and organize meetings in which invited experts can provide information to interested parties;

(c) The Initiative should establish capacity-building programmes, including through the provision of educational material, instrument distribution and/or access, national or regional expert centres, training of trainers, exchange programmes and competition and motivation programmes;

(d) The Initiative should serve as a catalyst for international collaboration by promoting the formation of common interest groups, conducting regular surveys of countries concerning their space competence profiles, developing a set of guidelines for collaboration, promoting multinational institutional agreements and creating regional expert centres;

(e) The Initiative should promote the exchange of knowledge and the sharing of data by raising awareness, promoting user-friendly mechanisms for data access and providing knowledge about self-supporting habitats for application, including for energy efficiency on Earth;

(f) Governments, institutions and individuals are encouraged to use space-based platforms for research in the following areas: psychology and social interaction in a multicultural, confined and isolated environment; vaccine development; nutritional, agricultural and food security; human physiology and aging; space technology for future exploration; and the space environment;

(g) Governments, institutions and individuals are encouraged to explore ground-based research for gravity-related science, preparing space-based experiments and making use of microgravity simulators (such as clinostats), microgravity instruments (such as parabolic flights, drop tubes and drop towers), hyper-gravity instruments (such as centrifuges) and software models;

(h) Governments, institutions and individuals are encouraged to explore the opportunities for commercial alternatives for educational and research activities in space, such as sub-orbital flights and long-duration experiments;

(i) Governments and institutions are encouraged to use space education as a tool for inspiring and motivating people and sustaining interest in science and technology; and

(j) Governments are encouraged to incorporate space education into the curricula of schools (in different subjects such as mathematics, physics, biology, chemistry and social science) and universities.

## III. ZERO-GRAVITY INSTRUMENT DISTRIBUTION PROJECT

UNOOSA is going to launch the Zero-gravity Instrument Distribution Project in response to the recommendations of the UN/Malaysia Expert Meeting on Human Space Technology in November 2011. In this project, zero-gravity instruments will be distributed to selected institutions such as schools, universities, and laboratories worldwide.

The main objectives of the project are to raise awareness and to build capacity in education and research in microgravity science, particularly in developing countries. An one-axis clinostat was selected for distribution because of the ease of use and potential scientific benefits of the project.

III.I One-axis clinostat
A clinostat is a laboratory instrument which can create a simulated microgravity condition by rotating samples around one or two axes, equalizing the gravity vector [14,15,16]. A one-axis clinostat has a horizontal rotational axis perpendicular to the gravity vector on the ground. Because of its simple structure and the ease of use, it is suitable for education and can also be utilized for research. This is the main reason why HSTI selected one-axis clinostats for distribution.

Figure 1 and Table 1 show a picture and the specifications, respectively, of the one-axis clinostat to be distributed. It has one rotational axis, the direction of which can be varied from 0 degrees (parallel to the ground) to 90 degrees (perpendicular to the ground). The rotational speed can be freely selected from 0 rpm to 90 rpm with a 0.5 rpm increment from 0 to 20 rpm and a 5 rpm increment from 20 rpm to 90 rpm. The rotational accuracy is 1%.

An one-axis clinostat has its own limitation if a sample put on the end of the rotational axis gets bigger and away from the axis. The two-dimensional rotation cannot effectively compensate for the gravity exerted on all parts of the sample body [17,18]. This difficulty, however, can be avoided by selecting small samples such as seeds or living cells.

There are several physical factors which degrade the performance of one-axis clinostats. The first factor is the angle between the rotational axis and the true horizontal plane [19]. If the angle is one degree, the axial residual acceleration is 0.02 g. It is important to set the rotational axis within 0.5 degrees from the true horizontal plane to reach $10^{-2}$ g. The second factor is the centrifugal force if a sample is placed away from the rotational axis [20]. A simple calculation shows that if a sample is placed one centimetre away from the rotational axis and the rotation speed is 10 rpm, the amount of the centrifugal force exerted on the sample is on the order of $10^{-3}$ g. There are other mechanical factors such as the non-constant rotational speed and the gyroscopic motion of the rotational axis which could be taken into account, in principle. However, it is expected that a well-designed experiment using a clinostat with a high mechanical quality should be able to achieve a $10^{-2}$-$10^{-3}$ g level of gravity unbalance.

Some of our target institutions for distribution are schools and universities. For teachers, HSTI is going to develop a teacher's guide which will be distributed with the clinostats. The teacher's guide will contain basic knowledge about microgravity, information on the clinostat, and samples of experiments as well as how to write experimental reports.

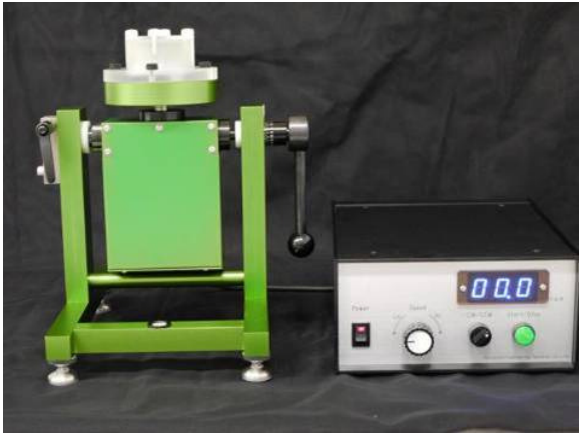

Figure 1: One-axis clinostat to be distributed

| Items | Specifications |
| --- | --- |
| Equipment size | Main body: 25 cm x 25 cm x 25 cm |
| | Control box: 23 cm x 20 cm x 11 cm |
| | Rotating portion: circular disk, 10cm in diameter |
| Number of rotational axes | One |
| Rotational speed | 0 - 90 rpm |
| | 0 - 20rpm: 0.5 rpm increment |
| | 20 - 90 rpm: 5 rpm increment |
| | Accuracy: 1% |
| Rotational axis angle | 0° (parallel to the ground)- |
| | 90° (perpendicular to the ground) |
| Rotational direction | CW or CCW |
| Input voltage | 100V - 240V |
| Building materials | Aluminium |

Table 1: Specifications of one-axis clinostat

III.II Project Outline

One cycle of the project is scheduled to last about two years, from the announcement of opportunity to the submission of the final report to UNOOSA. Currently, plans are to conduct two cycles by distributing 15 clinostats per cycle.

In order to select suitable institutions to receive the clinostats and increase the scientific value of the project, the HSTI Science Advisory Board (HSTI-SAB) will be established. HSTI-SAB will consist of several experts in the field of microgravity research, preferably having had research experience with clinostats. HSTI-SAB's primary task is to advise HSTI in the selection of institutions based on the evaluation of their educational or research proposals.

Table 2 shows the current timeline of the distribution project which has the following six phases:

A. Preparation
B. Announcement of Opportunity/Application
C. Selection
D. Distribution
E. Experiment
F. Reporting

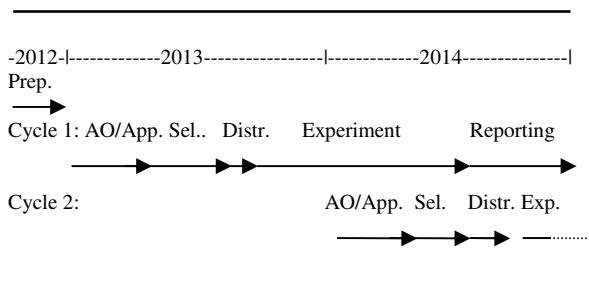

Table 2: Timeline of the distribution project

Phase A: During the preparation phase, UNOOSA will establish the HSTI-SAB, acquire 30 one-axis clinostats for distribution for two cycles, draft the announcement of opportunity and application form, develop the selection criteria, and conduct preliminary experiments.

Phase B: Announcement of Opportunity (AO) is scheduled to take place in January 2013 for Cycle 1 and in January 2014 for Cycle 2 and will be followed by a 3-month application period. The AO will be distributed through the United Nations networks to reach schools, universities, and laboratories worldwide.

Phase C: During the selection phase, all the applications will be reviewed and evaluated by predefined criteria and with the support of HSTI-SAB. After the final selection is finished, the selected institutions will be notified immediately. By that time, the teacher's guide will also be available for distribution.

Phase D: The clinostats along with the teacher's guides will be distributed to the selected institutions. A formal agreement between each selected institution and UNOOSA will be made to transfer the owner ship of the clinostat.

Phase E: The experiment phase will last about one year, during which period, institutions will use the clinostat to conduct experiments on the proposed projects using indigenous plant species.

Phase F: One of the conditions of the selection will be for the respective institutions to provide the annual reports on their activities with the clinostats to UNOOSA. HSTI will review annual reports from the selected institutions, consolidate all the reports, and publish the final report of each cycle of the distribution project.

Based on the scientific evaluation of the final reports submitted at the end of the $1^{st}$ and $2^{nd}$ cycles as well as the status of support from Member States, HSTI may extend the project to its $3^{rd}$ and $4^{th}$ cycles.

III.III Science with Clinostats

In order for humans to engage in long-term space travel, for example, a mission to Moon, Mars and beyond, we will have to advance space science and technology to allow us to grow vegetables and harvest food in space. Increasing the understanding of how organisms function in reduced gravity will give us a new understanding of fundamental biological processes. Studying how various plants, animals, and microorganisms interact with each other in a closed ecosystem in space is essential for developing advanced life support systems needed for long duration missions. These requirements are leading to a continued interest in microgravity science research.

The Earth is the host of more than 400,000 documented species of plant life [21]. By studying plants in microgravity, one can understand how plants respond to gravity. Since plants respond to gravity positively, and it is easy to obtain seeds in any regions and the seeds are small in size, plant seeds are one of the most suitable subjects for microgravity experiments with clinostats [22]. During Cycle 1 and Cycle 2 of the distribution project, the focus will be on observing the germination and early growth processes from seeds. The selected institutions will be requested to test at least 10 indigenous plant seeds on clinostats every year.

Although UNOOSA is going to evaluate the quality of the application proposals to select institutions to receive clinostats, an attempt will be made to distribute clinostats equally over every region on the Earth. This will ensure to be able to collect growth data for more than 400 different plant species by the end of Cycle 2.

The distribution project will provide opportunities for students and researchers from different regions to use their indigenous plant seeds to conduct microgravity educational and research activities. These efforts could lead to the establishment of a significant dataset on world plant growth with their gravity responses, which could be used to design future space experiments as well as contribute to the advancement of science.

IV. DISCUSSIONS

The Zero-gravity Instrument Distribution Project is one of the most challenging projects which the Programme on Space Applications has ever attempted. We would like to summarize some of the challenges we may encounter in the project and how to overcome these challenges.

In order for us to reach educational and research institutions around the world effectively, we would like to use all existing academic networks. One possible way is to use Regional Centres for Space Science and Technology Education Affiliated to the United Nations. The Regional Centres are educational institutions which provide 9-month post-graduate courses in space science and technology and are located in Morocco, Nigeria, Jordan, India, Mexico and Brazil. These Regional Centres have strong connections with academic institutions in neighbouring countries. By using the network of Regional Centres, we expect to reach educational and research institutions worldwide more quickly.

The key element for the success of the project is how well HSTI can achieve capacity building in microgravity education and research, particularly in developing countries. Distributing the clinostats and the teacher's guides are just the first steps in raising awareness and interests in this new field of science. How to sustain the interests will be the most challenging issue. The first solution for this challenge is to form a worldwide network on clinostat study with the selected institutions. With this network, all the participants in the project can communicate with each other, asking questions, answering them, and exchanging ideas. This will also strengthen their commitment to the project. The second solution is to form a group of scientists to support the project. Forming the HSTI-SAB is the first step in this endeavour. More scientists can participate in the project, and more effectively they can support the selected schools and universities in conducting experiments with the clinostats. This will also enhance the scientific value of the project. The third solution is to provide possible opportunities to conduct space experiments. It will depend heavily on what kind of scientific output the project can produce. Space-flight opportunities with indigenous plants would certainly attract the participation in this project from around the world.

## V. CONCLUSION

The United Nations Office for Outer Space Affairs is planning to conduct the Zero-gravity Instrument Distribution Project, which is part of the Human Space Technology Initiative (HSTI) within the framework of the United Nations Programme on Space Applications.

The project is to distribute one-axis clinostats which are expected to provide unique opportunities for students and researchers worldwide to observe the growth of indigenous plants in their countries in a simulated microgravity condition. A teacher's guide will also be distributed along with the clinostats to provide guided references.

The project is also expected to create a huge dataset of plant species with their gravity responses, which would be used to design future space experiments as well as contribute to the advancement of science.

## ACKNOWLEDGEMENT AND DISCLAIMER

We are grateful to all those who joined the activities of the HSTI, especially those who took part in the meetings and seminars in 2011 and 2012 and provided constructive comments to proceed with the HSTI implementation.

The views expressed herein are those of the authors and do not necessarily reflect the views of the United Nations.

## REFERENCES[1]

---

[1] *United Nations documents quoted in this paper are available from the website of the Office for Outer Space Affairs at http://www.unoosa.org and from the Official Document System of the United Nations at http://documents.un.org*